\documentclass[12pt,preprint]{aastex}

\usepackage{geometry}
\usepackage{graphicx}
\usepackage[symbol]{footmisc}
\usepackage{epstopdf}
\usepackage{subfigure}
\usepackage{amsmath}
\usepackage{float}

\def\kms{\rm\,km\;s^{-1}}
\def\half{\hbox{$\frac12$}}
\def\O{\mathcal O}
\def\H{\mathcal H}
\def\star{{\textrm{star}}}
\def\photon{{\textrm{\thinspace null}}}
\def\obs{{\textrm{obs}}}

\def\trav{{\textrm{trav}}}
\def\stwo{{\textrm{S2}}}

\def\static{{\textrm{static}}}
\def\kep{{\textrm{Kep}}}
\def\schw{{\textrm{Schw}}}
\def\fd{{\textrm{FD}}}
\def\mink{{\textrm{Mink}}}
\def\slo{{\textrm{S$_{\hbox{\tiny{LO}}}$}}}
\def\snlo{{\textrm{S$_{\hbox{\tiny{NLO}}}$}}}
\def\torq{{\textrm{torq}}}

\begin{document}

\title{Relativistic redshift effects \\ and the Galactic-center stars}

\author{Raymond Ang\'elil}
\and
\author{Prasenjit Saha}
\affil{Institute for Theoretical Physics, University of Z\"urich, \\
Winterthurerstrasse 190, CH-8057 Z\"urich, Switzerland}

\begin{abstract}
  The high pericenter velocities (up to a few percent of light) of the
  S stars around the Galactic-center black hole suggest that general
  relativistic effects may be detectable through the time variation of
  the redshift during pericenter passage.  Previous work has computed
  post-Newtonian perturbations to the stellar orbits.  We study the
  additional redshift effects due to perturbations of the light path
  (what one may call ``post-Minkowskian'' effects), a calculation that
  can be elegantly formulated as a
  boundary-value problem.  The post-Newtonian and post-Minkowskian
  redshift effects are comparable: both are $\O(\beta^3)$ and amount
  to a few $\kms$ at pericenter for the star S2.  On the other hand,
  the post-Minkowskian redshift contribution of spin is $\O(\beta^5)$
  and much smaller than the $\O(\beta^4)$ post-Newtonian effect, which
  would be $\sim0.1\kms$ for S2.
\end{abstract}

\pagebreak

\section{Introduction}

Over the past decade, dozens of fast-moving stars orbiting a large
compact mass (thought to be a supermassive black hole with mass
$\approx 4.4\cdot 10^6 M_\odot$) have been discovered
\citep{ghez1,gillessen}.  The highly eccentric orbits and the low
pericenter distances of these stars ($\sim 3\times 10^3$ of the
gravitational radius in the case of S2) provide a lucrative testing
ground for general relativistic perturbations to Keplerian
orbits. High resolution spectral and astrometric measurements of such
stars would also aid in the modeling of the mass distribution in the
Galactic center. Other consequences of general relativity, the
possible form of the metric itself (and the corresponding theories to
which the metric is a solution) could be studied.

The prospect for measuring general relativistic pericenter precession,
an $\O(\beta^2)$ effect (where $\beta$ is the pericenter velocity in
light units) for the Galactic-center stars is widely appreciated
\citep[see, for example,][]{jaroszynski, fragile} and is anticipated
to be observable by interferometric instruments currently under
development \citep{eisenhauer}.  Should stars further in be detected
in the future, precession effects at $\O(\beta^4)$ would become
measurable, enabling tests of no-hair theorems \citep{will08}.

Relativity also perturbs the kinematics of a
star. \cite{gillessenredshift} drew attention to the rapid velocity
changes that stars undergo around pericenter passage, and argued that
the $\O(\beta^2)$ effect of time dilation in the star's frame would be
measurable as a perturbation of the redshift.  \cite{saha} calculated
the $\O(\beta^3)$ kinematic contribution of space curvature $g_{ij}$
and $\O(\beta^4)$ effect of black hole spin $g_{0j}$, suggesting that
even the latter may become measurable with future interferometric
instruments. \cite{preto} presented a new orbit integration method in
the presence of space curvature, spin, and Newtonian Galactic
perturbations.

In this paper, we extend previous work to include the redshift
contributions that come from the effect of space curvature and spin on
the light path between the star and the observer.  Specifically, we
will compute the redshift of a moving point source in a weak-field
approximation of a Kerr metric, identify the various contributions,
and investigate the scaling of these signals with orbital size.

We formulate the problem as two photons emitted from a stellar orbit,
an infinitesimal proper time interval $\Delta \tau$ apart.  Both
photons hit the observer, but with a difference $\Delta t$ in arrival
time.  The redshift $z$ is then
\begin{equation}
\Delta t = (1+z) \Delta \tau.
\label{zdef}
\end{equation}
The physical process is the same as in the calculation of the spectrum
of an accretion disk \citep[e.g.,][]{mueller}, only the computational
strategy needed is different.  In the accretion-disk case, photons are
shot backwards from the observer in a range of directions, but in the
general direction of the extended source (the disk). By considering
the collision events of each ray with the source surface, the
quantities of interest may be read off. In the stellar-source case,
however, it is necessary to solve for the initial direction of a
photon so that it will reach the observer. Hence we have a
boundary-value problem.

\begin{figure}[!ht]
\begin{center}
\subfigure{\includegraphics[height=2.4in, width = 1.6in, clip = true, trim =
  50pt 50pt 115pt 95pt]{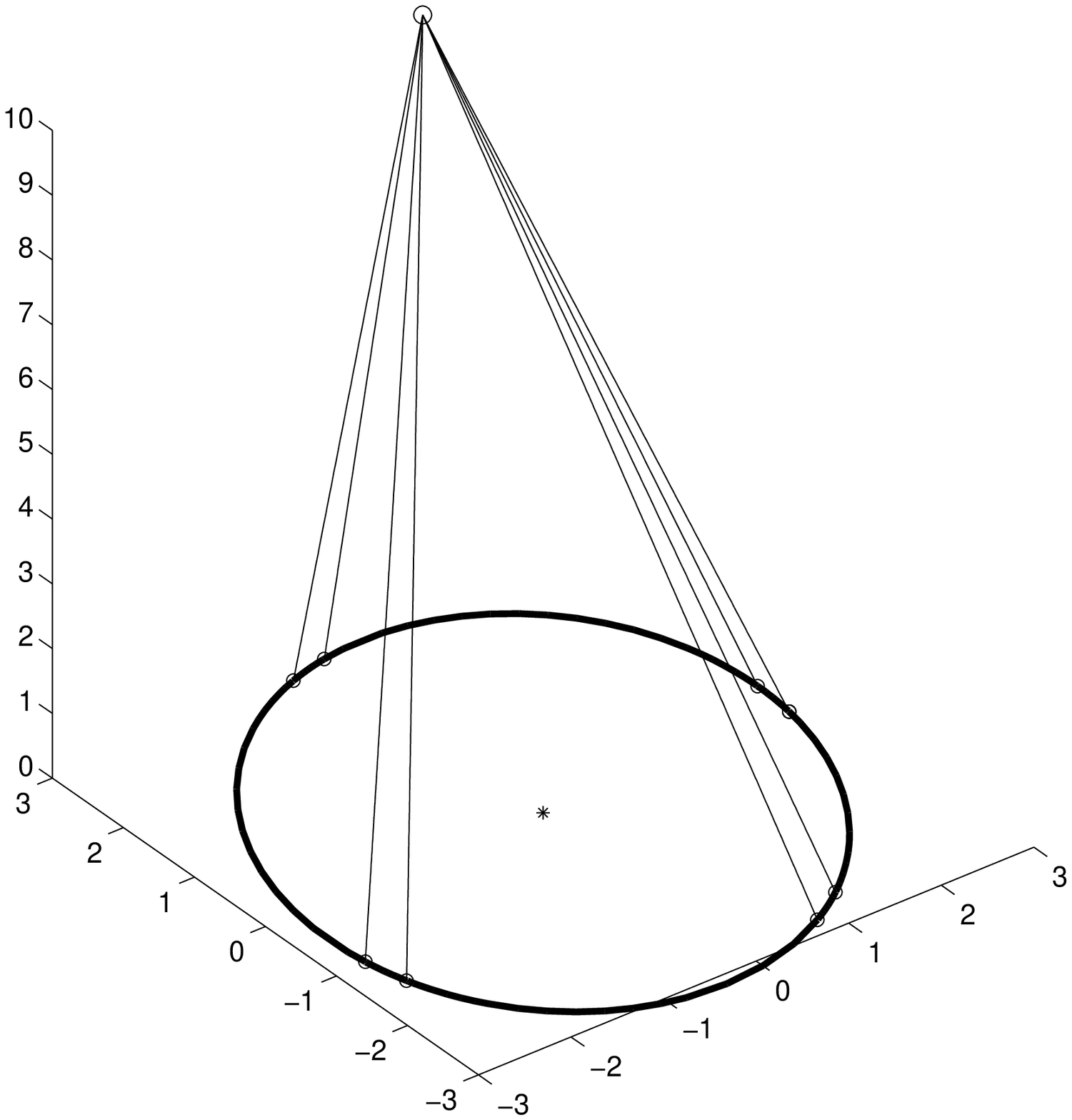}}
\subfigure{\includegraphics[height=2.4in, width = 1.6in, clip = true, trim =
  50pt 50pt 115pt 95pt]{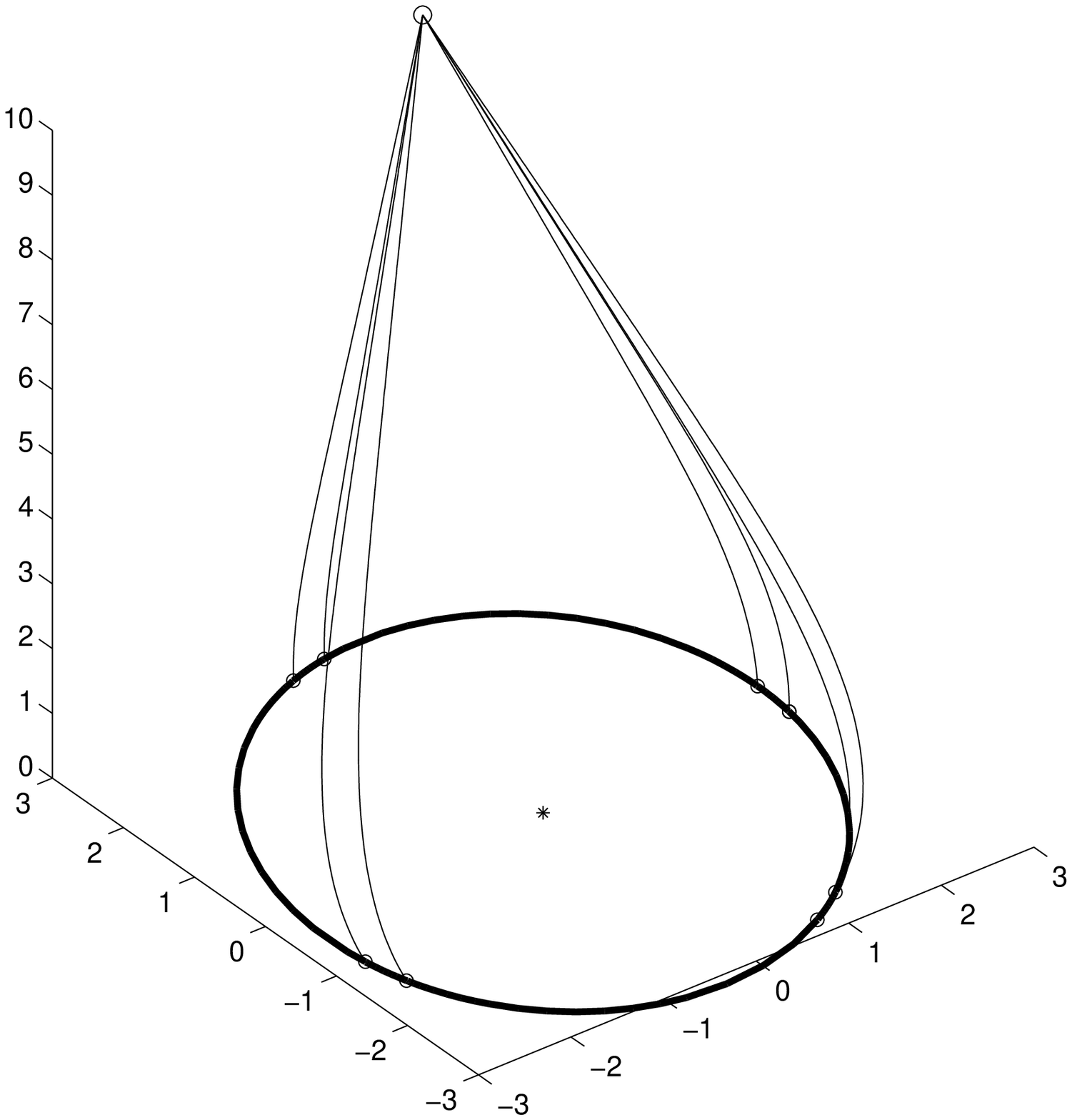}}
\subfigure{\includegraphics[height=2.4in, width = 1.6in, clip = true,
  trim = 50pt 50pt 115pt 95pt]{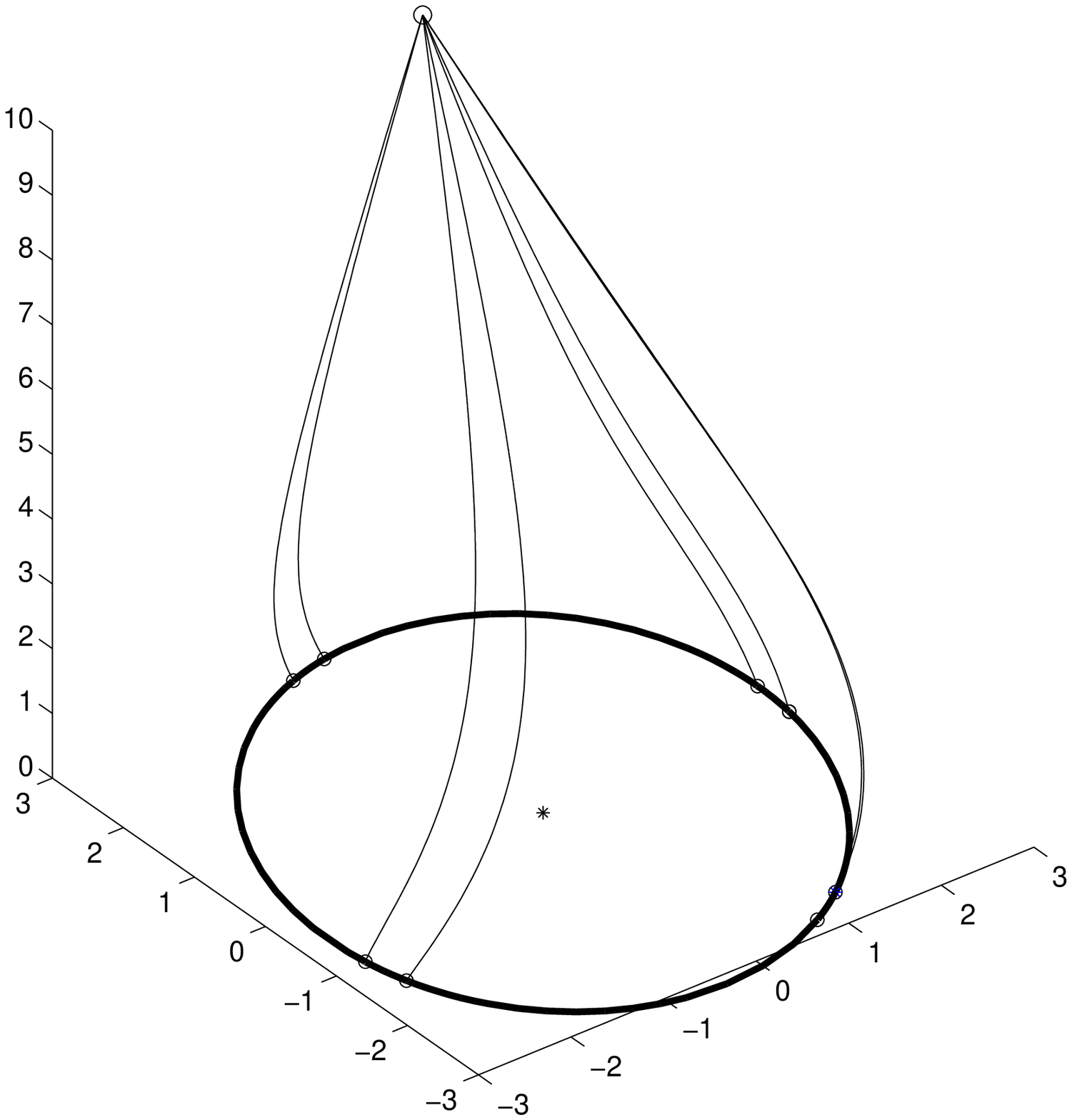}}
\end{center}
\caption{Schematic illustration of the method.  Pairs of photons are
  emitted at slightly different proper times along the orbit, in
  precisely the right direction to reach the observer.  Finding these
  photons is a boundary-value problem, and once found, each photon
  pair allows us to calculate the redshift at that point on the orbit
  by evaluating (\ref{daProblem}).  At the left of the figure we have
  Minkowski photons, which move in straight lines.  In the middle we
  have Schwarzschild photons, which are lensed.  On the right, we have
  frame-dragged photons. The time difference between the emission of each photon in a pair has been exaggerated here for visual clarity. Note that it is only the star's unrealistic proximity to the black hole 
  that allows for such a visible depiction of the different effects.} 
  
\label{example}
\end{figure}

Figure \ref{example} is a schematic illustration of our method.
Photons are emitted from nearby spacetime points on the stellar orbits
and travel to the observer.  In the left-hand picture, the star emits
``Minkowski photons'' which feel no space curvature and travel in
straight lines; redshift depends only on the velocity and time
dilation of the star.  This is in effect the approximation used in
previous work.  In the middle picture, the star emits ``Schwarzschild
photons'' which feel space curvature. In the right-hand picture, the
star emits ``frame dragging photons'' which feel spin as well as space
curvature.

Below, \S~\ref{Algorithm} details the problem to be solved and the
method used for calculation of the redshift. The Matlab scripts
implementing our algorithm are available as an online supplement. Then
\S~\ref{theMetric} presents the black-hole model and associated
metrics which we use in our approach, and \S~\ref{scalesec} derives
how the various effects scale with orbit size.  We apply our algorithm
to the star S2, and detail the results in Section~\ref{Results}.

\section{The redshift-calculation method}

\subsection{A boundary-value problem}\label{Algorithm}

In order to calculate the redshift of a moving star as observed by a
fixed observer, we need to solve the geodesic equations for both the
star and for photons.  Geodesic equations are commonly expressed in
terms of the Lagrangian,
\begin{equation}
\mathcal{L} = \half g_{\mu\nu} \dot x^\mu \dot x^\nu
\end{equation}
with dots denoting derivatives with respect to the affine parameter.
But an equivalent formulation exists in terms of a Hamiltonian
\begin{equation}
\mathcal{H} = \half g^{\mu\nu} p_\mu p_\nu.
\end{equation}
We will follow the latter in this work.  Numerically $\mathcal H =
\mathcal L$.  We will in fact employ two Hamiltonians\footnote{In an
  effort to keep the distinctions between effects on the star orbit
  and those on the light path as perspicuous as possible, we adopt
  subscripts for orbital effects and superscripts for light-path
  effects. This notation may be found on Hamiltonians $\H$ and
  redshifts $z$, and has nothing to do with covariant/contravariant
  indices.}  $\H_\star$ and $\H^\photon$, which are really two
different approximations for the
same Hamiltonian. Numerically $\H^\photon=0$ of course.  We will
write $\lambda$ for the affine parameter of the star, and $\sigma$ for
that of a photon.

Consider two photon trajectories, emitted at two points on the star's
orbit $\Delta\lambda$ apart in the affine parameter, with both photons
terminating at the observer and arriving at time $\Delta t$ apart in
the observer's frame.  Since $d\tau = \sqrt{|g_{\mu\nu}dx^\mu dx^\nu|}$,
we are able to express the proper time between the emission events in
terms of the affine parameter as
\begin{equation}
\Delta \tau =\Delta \lambda\sqrt{|2\H_\star|}
\end{equation}
where $\H_\star$ is evaluated at either point of emission.  Comparing
with the definition~(\ref{zdef}) of the redshift, we have
\begin{equation}\label{daProblem}
z = \frac{\Delta t}{\Delta \lambda\sqrt{|2\H_\star|}} - 1 .
\end{equation}

We now need to compute $\Delta t$ for a given $\Delta\lambda$.  To do
this, we begin by calculating the orbit of the star for some chosen
initial conditions, by solving Hamilton's equations for $\H_\star$
with $\lambda$ as the independent variable. The temporal component of
the generalized momentum is set at $p_t=-1$. This amounts to choosing
the units for $\lambda$ such that $dt=d\lambda$ outside of
gravitational fields.  We then choose a point on the star's orbit
whose observed redshift we wish to calculate, and from this point we
seek a photon that will reach the observer.

Consider a function $\Phi^j$, which effectively shoots a photon by
integrating Hamilton's equations for $\H^\photon$ with given initial
conditions at affine parameter $\sigma = 0$ and returns the 3-position
at $\sigma = 1$.  We write
\begin{equation}
\Phi^j\left(t, x^i, p_i\right) = x^j|_{\sigma=1}
\end{equation}
where $i,j = 1,2,3$ and the initial $p_t$ is chosen such that
$\H^\photon=0$.  We pass the function $\Phi^j$ to a root-finder, and
solve for the root of
\begin{equation}\label{solvethis}
f\left(p_i\right) = \Phi^j - x^j_\obs
\end{equation}
by varying the initial 3-momentum $p_i$.  Naturally, we may not adjust
the $x^i$, as we are interested in a specific point on the star's
orbit.

The root-finding algorithm requires a set of initial guesses for the
initial $p_i$ of the photon.  On this account, we shoot an
initial-guess photon from the star position in the direction of the
observer, that is, we evaluate $\Phi^j$ with trivial initial
conditions.  These initial-guess values for the $p_i$ shoot in the
direction of the observer ignoring curvature. However, because the
spacetime is indeed curved, this photon will {\em not\/} hit the
observer. It serves only to start the root-finding algorithm.

Once the root-finder reports a solution within specified tolerance
level, we move the star a very short distance ($\Delta\lambda$ in
affine parameter) along its orbit, and repeat the above procedure, now
solving for the sought-after trajectory at the star's new
position. The difference in arrival times of the two photons is
$\Delta t$, which we insert into (\ref{daProblem}) to evalate the redshift.
   
Thus far we have solved for two photon trajectories. This computation
has enabled us to calculate the redshift at a chosen point on the
orbit. We may then repeat this process at further points along the
orbit, garnering results as much as the required resolution demands.

\subsection{Post-Newtonian and post-Minkowskian approximations}
\label{theMetric}

The spacetime outside a spinning black hole is described by the Kerr
metric.  Since the Galactic-center stars are far from the horizon,
approximations valid only at large $r$ can be used to study general
relativistic effects.  Accordingly, we first derive two perturbative
Hamiltonians, a post-Newtonian $\H_\star$ for stars, and a
post-Minkowskian\footnote{The term `post-Minkowskian' is often used as
a synonym for `weak-field metric'.  We are using it, however, to refer
specifically to light paths that deviate slightly from special relativity.}  approximation
$\H^\photon$ for photons.  Different physical effects come into play
at different orders of the perturbation parameter\footnote{In this
paper $\epsilon,\epsilon^2$ and so on are just labels to keep track of
different orders. Numerically $\epsilon=1$.}  $\epsilon$, and we will
show these numerically by toggling different terms on and off.

Taking the Kerr metric in Boyer-Lindquist coordinates with geometric
units $GM = c = 1$ (leaving us with a unit of length equal to the
gravitational radius $GM/c^2 \simeq 5 \times 10^6$ km for the
Galactic-center black hole) we have the full Hamiltonian
\begin{eqnarray}\label{kerrmetric}
\H_{\textrm{Kerr}} &=&
  \frac{\left(r^2+s^2\right)^2-s^2\Delta \sin^2\theta}{2\rho^2\Delta}p_t^2
- \frac{\Delta}{2\rho^2}p_r^2 - \frac{1}{2\rho^2}p_\theta^2  \nonumber \\
\noalign{\medskip}
& & 
- \frac{\Delta-s^2\sin^2\theta}{2\Delta \rho^2\sin^2\theta}p_\phi^2
+ \frac{2sr}{2\rho^2\Delta} p_t p_\phi  \\
\end{eqnarray}
where
\begin{equation}
\Delta\equiv r^2-2r+s^2
\quad\textrm{and}\quad
\rho^2\equiv r^2+s^2\cos^2\theta .
\end{equation}
and $s$ denotes the black hole spin parameter.

Let us first consider the dynamics of the star.  Sufficiently far from
the black hole, the post-Newtonian approximation
\begin{equation}\label{velocity}
v^2 \sim 1/r
\end{equation}
applies.  If we choose $v \sim \O\left(\epsilon\right)$ then
by (\ref{velocity}) $r$ is
$\O\left(\epsilon^{-2}\right)$. Correspondingly, we force the
velocity terms $p_r$, $p_\theta/r$ and $p_\phi/r$ to be
$\O\left(\epsilon\right)$. Making the following
replacements in (\ref{kerrmetric})
\begin{equation}
r\rightarrow\epsilon^{-2}r, \quad
p_r\rightarrow\epsilon p_r, \quad
p_\theta\rightarrow\epsilon^{-1} p_\theta \quad \mathrm{and} \quad
p_\phi\rightarrow\epsilon^{-1} p_\phi
\end{equation}
and keeping terms to $\O\left(\epsilon^5\right)$, we obtain
\begin{equation}\label{timelikeMetric}
\H_{star} = -\frac{p_t^2}{2}+
\left(\frac{p_r^2}{2} + \frac{p_\theta^2}{2r^2}+ 
\frac{p_\phi^2}{2r^2\sin^2\theta}-\frac{p_t^2}{r}\right) \epsilon^2 
- \left(\frac{2p_t^2}{r^2}+\frac{p_r^2}{r}\right) \epsilon^4 
- \frac{2s p_t p_\phi}{r^3} \, \epsilon^5 .
\end{equation}
We can abbreviate (\ref{timelikeMetric}) as
\begin{equation}\label{Hstar}
\H_\star = \H_\static + \epsilon^2\, \H_\kep + \epsilon^4\, \H_\schw
           + \epsilon^5\, \H_\fd .
\end{equation}
Here $\H_\static$ produces motionless geodesics, $\H_\kep$ gives the
Keplerian phenomenology, $\H_\schw$ is the weak-field Schwarzschild
contribution that produces pericenter precession, while the
angular-temporal term $\H_\fd$ produces the Lens-Thirring effect or
frame dragging.

Continuing now to photon trajectories, we remark that by the
equivalence principle, the full Hamiltonian is exactly the same for
photons and stars.  However, the same terms can have different orders
in the two regimes, prompting approximation $\H^\photon$.
In particular, the approximation (\ref{velocity}) obviously
does not hold for photons, which have $v^2=1$,
implying the scalings
\begin{equation}
r \rightarrow \epsilon^{-2}r, \quad
p_\theta\rightarrow \epsilon^{-2} p_\theta \quad \mathrm{and} \quad
p_\phi\rightarrow\epsilon^{-2} p_\phi
\end{equation}
with $p_r$ being $\O(1)$.
Expanding as before, we obtain
\begin{align}
\H^\photon = & -\frac{p_t^2}{2}+\frac{p_r^2}{2}+\frac{p_\theta^2}{2r^2}+
\frac{p_\phi^2}{2r^2\sin^2\theta}
-\left(\frac{p_t^2}{r}+\frac{p_r^2}{r}\right)\epsilon^2+\nonumber\\
&-\left(
\frac{2p_t^2}{r^2}+\frac{2sp_tp_\phi}{r^3}
-\frac{s^2\sin^2\theta}{2r^2}p_r^2
+\frac{s^2\cos^2\theta}{2r^4}p_\theta^2
\right)\epsilon^4
\label{nullmetric}
\end{align}
- a different selection of terms compared to the post-Newtonian case.
There is no term at $\O(\epsilon^5)$.  We may abbreviate
(\ref{nullmetric}) as
\begin{equation}\label{Hphoton}
\H^\photon = \H^\mink + \epsilon^2\, \H^\slo +
\epsilon^4 \left( \H^\snlo + \H^\fd + \H^\torq \right) .
\end{equation}
At zeroth order we have special relativistic or Minkowski photons.
The leading order Schwarzschild effect $\H^\slo$ gives the
gravitational deflection of light.  At
$\O\left(\epsilon^4\right)$, however, there are three
distinct effects: first $\H^\snlo$ gives a next-to-leading order
correction to the Schwarzschild effect, the off-diagonal term $\H^\fd$
gives frame dragging again but for photon trajectories, while
$\H^\torq$ provides a torque proportional to $s^2$.

\subsection{Pseudo-cartesian coordinates}

The spatial Boyer-Lindquist coordinates $r,\theta,\phi$ are
convenient for computing stellar orbits, but not well suited for
photon paths.  The photon paths are nearly straight lines, but since
the observer is much further from the black hole than the source, tiny
variations in $\theta$ and $\phi$ at the observer imply large
distances.  As a result, both the integrator and the root-finder become
susceptible to roundoff error.

To cure the problem, we change to pseudo cartesian coordinates. They are not purely cartesian, as the 
surface $x^2 + y^2 + z^2$ is not spherical.
\begin{equation}
x = r\sin\theta\cos\phi, \quad y = r\sin\theta\cos\phi, \quad
z = r\cos\theta
\end{equation}
The corresponding momenta are readily derived by completing the
canonical transformation, leading to the usual relations
\begin{equation}
p_r = \frac{\textbf{x}\cdot\textbf{p}}{r}, \quad
p_\phi = \left(\textbf{x}\times \textbf{p}\right)_z, \quad
p_r^2 +\frac{p_\theta^2}{r^2} + \frac{p_\phi^2}{r^2\sin^2\theta}
= \textbf{p}^2.
\end{equation}
The form of $\H^\photon$ changes accordingly. The Minkowski part bears the
familiar form
\begin{equation}
\H^\mink = -\frac{p_t^2}{2} +  \frac{\textbf{p}^2}{2}. 
\end{equation}
The leading-order Schwarzschild terms are
\begin{equation}
\H^\slo = -\frac{p_t^2}{r}
- \frac{\left(\textbf{x}\cdot \textbf{p}\right)^2}{r^3},
\end{equation}
and the associated $\O(\epsilon^4)$ term retains its previous
form of
\begin{equation}\label{NLO}
\H^\snlo = -\frac{2p_t^2}{r^2}.
\end{equation}
Also at $\O(\epsilon^4)$ we have the frame-dragging term
\begin{equation}\label{FD}
\H^\fd =  - \frac{2 s p_t}{r^3}\left(\textbf{x}\times\textbf{p}\right)_z,
\end{equation}
and the torquing terms
\begin{equation}\label{torque}
\H^\torq = \frac{s^2}{2r^2}\left(1-\frac{z^2}{r^2}\right)
\left(\frac{\textbf{x}\cdot \textbf{p}}{r}\right)^2
- \frac{s^2}{2z^2}\left[\textbf{p}^2
- \left(\frac{\textbf{x}\cdot \textbf{p}}{r}\right)^2
- \frac{\left(\textbf{x}\times\textbf{p}\right)^2_z}{r^2-z^2}\right] .
\end{equation}
The torquing terms are the most difficult to integrate
numerically. This is because while the terms themselves are at
$\O\left(\epsilon^4\right)$, they involve quotients of particularly
high powers. Taking derivatives of $\H^\torq$ further bloats these
bottom and top heavy fractions, and provokes roundoff errors.  We
argue below that the torquing terms are in any case unimportant for
the known Galactic-center stars.  Hence we omit these terms in our
numerical work.

\subsection{Scaling properties of redshift contributions}\label{scalesec}

We can infer the scaling with orbital size $a$ of the redshift
contributions $\Delta z$ of the various perturbative terms in
$\H_\star$ with the following deliberation. Consider a perturbative
term
\begin{equation}
\Delta \H_\star \sim a^{-n}
\end{equation}
Since $\H_\star$ is constant along the orbit, any variation in
$\Delta\H_\star$ has to be balanced by a variation in the unperturbed
Hamiltonian. Since the latter scales as $\H_\star \sim 1/a $, we have
$\Delta \H_{star} \sim \Delta a/a^2$, giving
\begin{equation}
\Delta a \sim a^{2-n}
\end{equation}
Furthermore, since the orbital velocity scales as $v \sim a^{-1/2}$,
we have $\Delta v \sim a^{-3/2}\,\Delta a$ and hence the redshift
signal
\begin{equation}\label{dzstar}
\Delta z \left(\Delta H_\star\right)\sim a^{\frac12-n} .
\end{equation}

A similar argument can be made for the perturbative terms
$\Delta\H^\photon$.  In this case we compare photons emitted from the
same point, only with different Hamiltonians. While the redshift
signal from the previous case necessitated our consideration of the
stellar velocity only, in analyzing the gravitational redshift
signal, we are naturally interested in the time difference $\Delta t$
due to $\Delta H$. Let the perturbation be
\begin{equation}
\Delta \H^\photon \sim r^{-n}.
\end{equation}
Since the light travel time is an integrated quantity, we expect the
change in the light travel time to scale as
\begin{equation}
\Delta t_\trav \sim r^{1-n}
\end{equation}
The change $\Delta t_\trav$ in light travel time  must not be confused
with the difference $\Delta t$ in arrival time of
two photons.  For the latter quantity we get $\Delta t \sim \Delta r/r^n$ and since $\Delta r$ between two photons is $\sim v \sim a^{-1/2}$
we derive
\begin{equation}\label{dzphoton}
\Delta z \left(\Delta \H^\photon\right) \sim a^{-\frac12-n} .
\end{equation}

Using (\ref{dzstar}) and (\ref{dzphoton}) and recalling that each
power of $\epsilon$ in the Hamiltonians represents a scaling factor of
$r^{-1/2}$ we can read off the following scalings.
\begin{equation}\label{scalings}
\left(
\begin{array}{ccc}
& \H_\schw \\
& \H_\fd   \\
& \H^\slo  \\
\H^\snlo & \H^\torq & \H^\fd
\end{array}
\right)
\Rightarrow \Delta z \propto
\left(
\begin{array}{l}
a^{-3/2} \\
a^{-2}   \\
a^{-3/2} \\
a^{-5/2}
\end{array}
\right)
\end{equation}

\section{Application to S2-like orbits}\label{Results}

We select S2 for a case study, since S2 has the shortest orbit and one
of the highest pericenter velocities of all the known stars orbiting
Sagittarius A$^*$, and hence provides us with perhaps the best
opportunity to observe general-relativistic effects.

\begin{figure}[!ht]
   \subfigure{\includegraphics[height = 2.5in, clip = true, trim = 0 0 20pt 0]{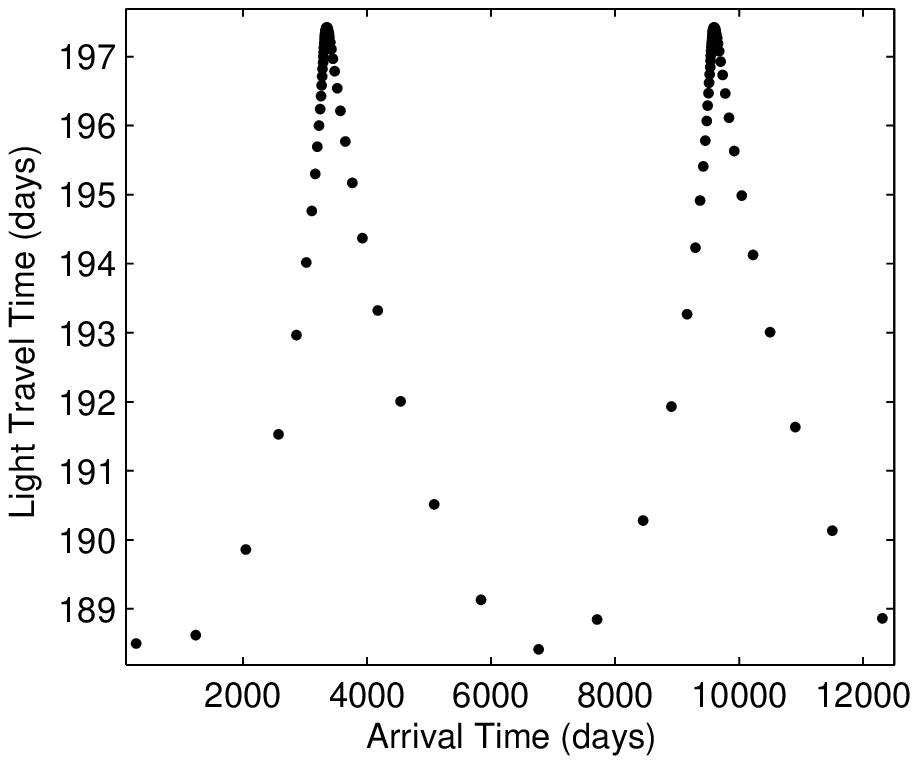}}
   \subfigure{\includegraphics[height = 2.5in, clip = true, trim = 0 0 20pt 0]{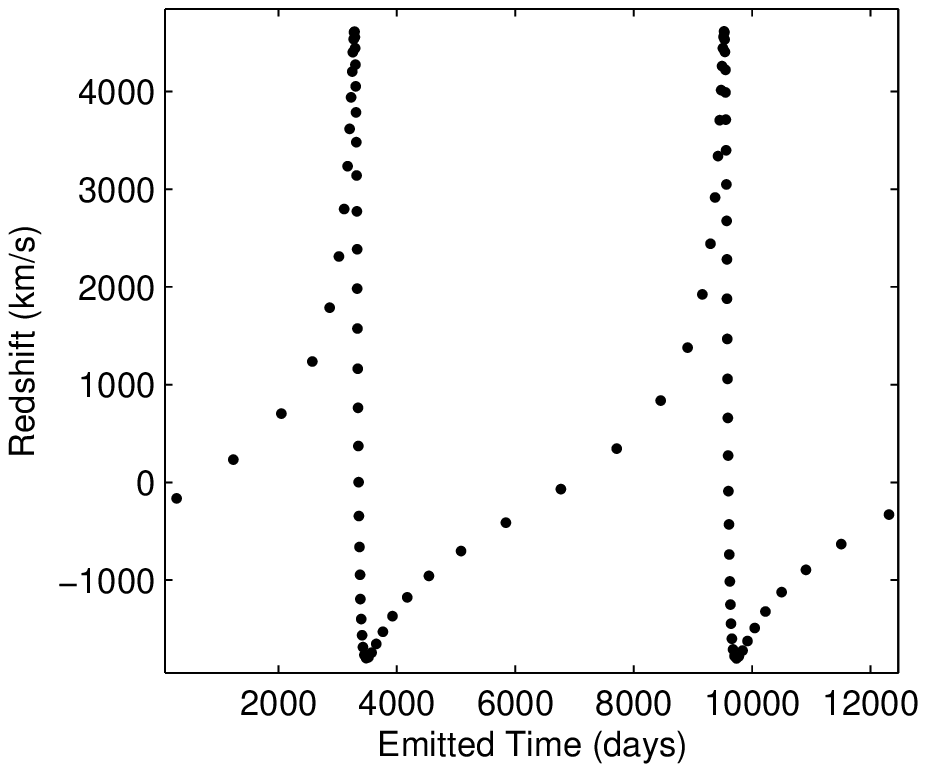}}
  \caption{A redshift calculation for an S2-like star over two orbits.
    The left panel shows the time each photon takes to travel from the
    star to the observer. The star begins at apocenter, which happens
    to be closer to the observer than the pericenter. Therefore the
    light travel time naturally increases as the star moves towards
    its pericenter. The right panel shows the redshift. The peaks
    occur during pericenter passage due to the high pericenter passage
    velocities.  The calculation includes post-Newtonian and
    post-Minkowskian effects, but these are too small to see in this
    figure.}
  \label{zfig}
\end{figure}

Figure~\ref{zfig} shows a redshift calculation for a star with S2's orbital parameters. These are taken from \cite{gillessen}. The gravitational radius is taken as $5\times10^6\;\rm km$. For definiteness, we take the spin to be maximal, and pointing towards Galactic North. Disc seismology models \citep{spin} put the spin at $s \approx 0.44$. The direction however, remains unknown.  

All the contributions to $\H_\star$ in
(\ref{Hstar}) are included for the orbit calculation.  The photon
trajectories include all contributions to $\H^\photon$ in
(\ref{Hphoton}) except for $\H^\torq$.

Naturally we would like to compute the redshift contributions of the
various relativistic terms, and verify that they follow the expected
scalings (\ref{scalings}).  In order to do this, we examine the
differences between redshifts computed from different post-Newtonian
and post-Minkowskian cases.  This allows us to isolate the effects of
$\H_\schw,$ $\H_\fd,$ $\H^\slo$ and $\H^\snlo+\H^\fd$, as follows.

\begin{figure}[!ht]
\begin{center}
\includegraphics[height = 4.0in, clip = true, trim = 20pt 0 0 0]{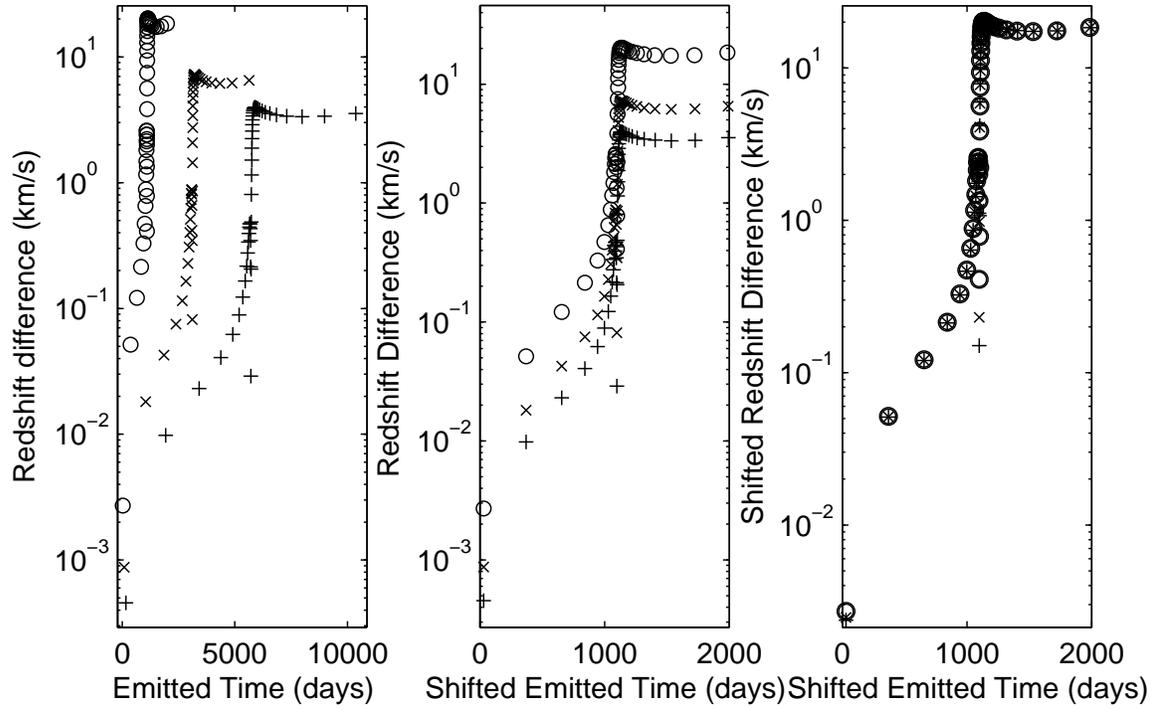}
\end{center}
\caption{Redshift difference $z^\mink_\schw-z^\mink_\kep$ showing the
  contribution of $\H_\schw$. The left panel shows the redshift
  difference for three orbits, one with the parameters of S2, and the
  other two with $a$ doubled or halved.  In the middle panel, the
  horizontal scale is stretched $(a/a_\stwo)^{-3/2}$.  In the right
  panel the redshift difference is scaled by $(a/a_\stwo)^{-3/2}$.}
\label{case1}
\end{figure}

\begin{figure}[!ht]
\begin{center}
\includegraphics[height = 4.0in,clip = true, trim = 20pt 0 0 0]{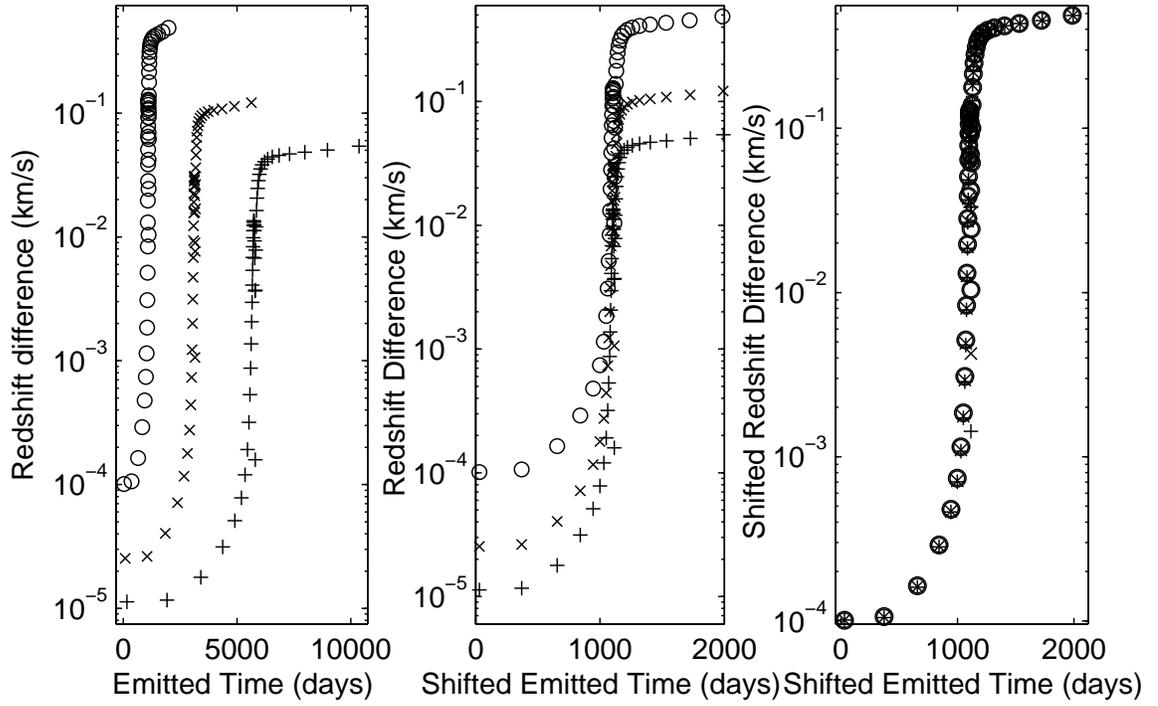}
\end{center}
\caption{Redshift difference $z^\mink_\fd-z^\mink_\schw$ showing the
  contribution of $\H_\fd$. The scheme follows Figure~\ref{case1}
  except that the redshift difference is scaled by
  $(a/a_\stwo)^{-2}$.}
\label{case2}
\end{figure}

\begin{figure}[!ht]
\begin{center}
\includegraphics[height = 4.0in, clip = true, trim = 20pt 0 0 0]{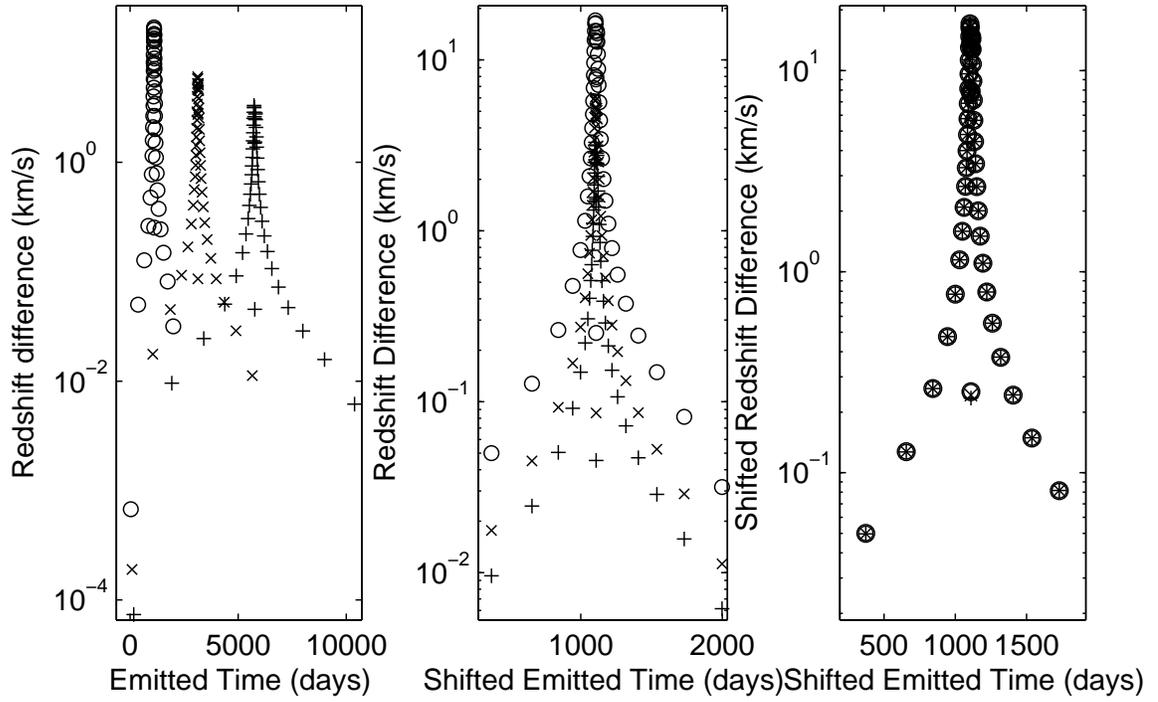}
\end{center}
\caption{Redshift difference $z^\slo_\schw-z^\mink_\schw$ showing the
  contribution of $\H^\slo$. The scheme follows Figure~\ref{case1},
  the redshift difference being scaled again $(a/a_\stwo)^{-3/2}$.}
\label{case3}
\end{figure}

\begin{figure}[!ht]
\begin{center}
 \includegraphics[height = 4.0in, clip = true, trim = 20pt 0 0 0]{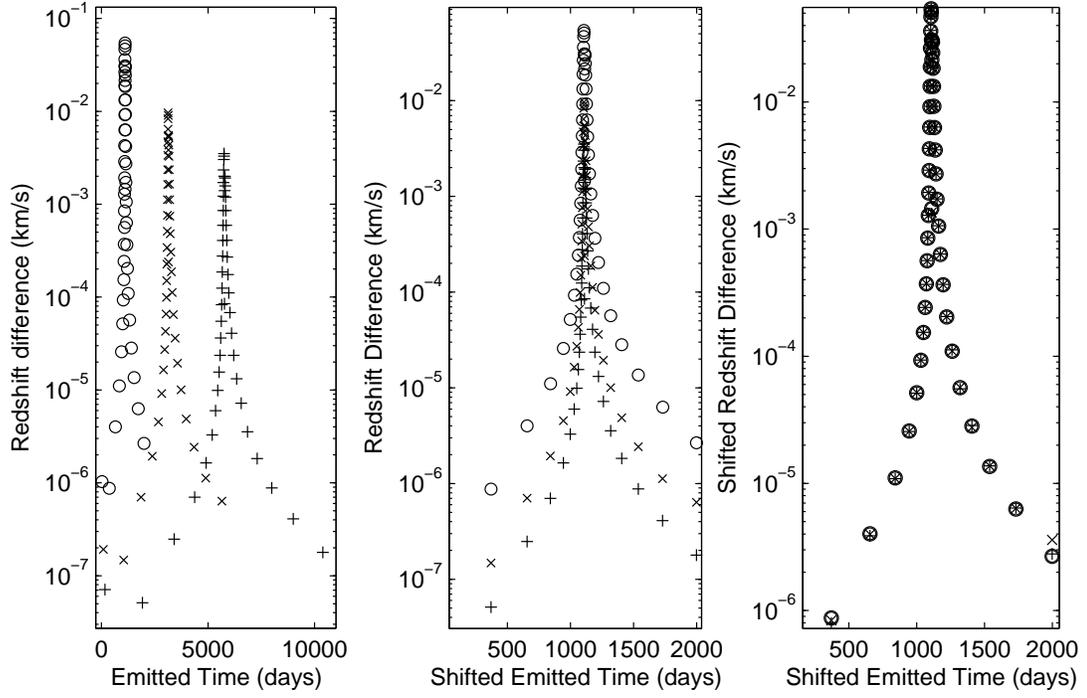}
\end{center}
\caption{Redshift difference $z^\fd_\fd-z^\slo_\fd$ showing the
  contribution of $\H^\snlo+\H^\fd$. The scheme follows
  Figures~\ref{case1}--\ref{case3}, with the redshift difference
  scaled by $(a/a_\stwo)^{-5/2}$. Here we have taken the spin as maximal, $s = 1$. This signal is exactly proportional to s.}
\label{case4}
\end{figure}

\begin{enumerate}
\item To isolate $\H_\schw$ we compute the redshift difference
  $z_\schw^\mink-z_\kep^\mink$.  By $z_\schw^\mink$ we mean that the
  star is followed using terms in $\H_\star$ up to $\H_\schw$ and the
  photons are followed using $\H^\photon$ up to $\H^\mink$.  The same
  naming convention applies to $z_\kep^\mink$ and to other expressions
  of this type below.

  Figure~\ref{case1} shows the redshift difference, calculated for
  three orbits going from apocenter to apocenter.  One orbit has the
  parameters of S2; the two others have $a=2a_\stwo$ and $\half
  a_\stwo$ with the other orbital parameters being the same.  The
  redshift difference increases till pericenter and then declines
  somewhat, but not to its previous apocentric value, because the
  relativistic orbit experiences prograde Schwarzschild precession
  whereas the Keplerian orbit does not, and the resulting phase change
  in the orbit gives an increasing contribution to the redshift.

  For S2 parameters, the maximum redshift contribution of $\H_\schw$
  is found to to be $\simeq 7\kms$.  For the other two stars, upon
  rescaling the redshift differences by $(a/a_\stwo)^{-3/2}$ and the
  orbital time also by $(a/a_\stwo)^{-3/2}$, the results can be
  overlaid almost perfectly on those of the S2-like star.

\item To isolate $\H_\fd$ we then compute the redshift difference
  $z_\fd^\mink-z_\schw^\mink$.  For an S2-like orbit the signal is
  around $0.1\kms$ at pericenter, and as Figure~\ref{case2} shows, the
  signal scales as $a^{-2}$.

\item To isolate $\H^\slo$ we compute $z^\slo_\schw-z^\mink_\schw$ and
  illustrate this difference in Figure~\ref{case3}.  There is no precession-related
  redshift effect involved, because the photon types being compared
  refer to the same stellar orbits.  The signal scales as $a^{-3/2}$
  and for the S2-like orbit the maximum is $\simeq2\kms$.

  We see that the Schwarzschild terms in the stellar orbit and in the
  light path give comparable contributions to the redshift.  To detect
  the Schwarzschild effect, it is necessary to take both into account.

\item Finally, we isolate $\H^\snlo+\H^\fd$ by computing
  $z_\fd^\fd-z_\fd^\slo$.  Figure~\ref{case4} verifies the expected
  $a^{-5/2}$ scaling and shows that the maximum signal for S2
  parameters is $\sim10^{-2}\kms$.  Thus we see that the
  frame-dragging on photons is much smaller than on stars.  Similarly,
  $\H^\snlo$ makes a much smaller contribution than $\H_\fd$.  We
  expect the contribution of $\H^\torq$ would be similarly small,
  though we have not calculated it.

  We see that in order to measure the leading-order frame-dragging
  effect on Galactic-center stars, it is sufficient to consider
  Schwarzschild photons.

\end{enumerate}

Of course, the computation method for weak redshift signals contains
numerical errors, especially for higher-order effects being evaluated
at large distances from the black hole.  The numerical noise in our
implementation drowns out the the leading-order Schwarzschild effect
for scaled S2-like orbits with $a \approx 3\cdot10^8$ --- at which
point the pericenter redshift signal is $\approx 5\cdot10^{-6}\kms$.
Numerical noise would overwhelm the frame-dragging signal for scaled
S2-like orbits with at $a\approx 1.2\cdot 10^5$.  This being an order
of magnitude larger than $a_{S2}$, and so we are in good shape to
calculate the frame-dragging redshift contribution. For redshift
signal contributions beyond those of frame-dragging, the numerical
noise in our Matlab implementation of the algorithm is intolerable for
$a_{S2}$. Were we calculating these effects for stars closer to the
SBH, the higher-order signals would be stronger, and therefore less
prone to round-off.  In summary, for the known Galactic-center stars,
the numerical noise in calculating relativistic redshifts will be well
within the observational errors, even with the next generation
spectral instrumentation.

\section{Summary and Outlook}

Some stars in orbit around the Galactic-center black hole reach
velocities of a few percent light at pericenter, and the time-varying
redshift of these stars during pericenter passage has small but
distinctive perturbations from general relativity.

The redshift is dominated by the line of sight velocity, which for the
star S2 reaches $v\sim5\times10^3\kms$ at pericenter.  The leading
perturbations are from time dilation because (a)~the star is moving,
and (b)~because it is in a potential well.  Both of these make
$\O(\beta^2)$ contributions to the redshift (where $v$ is the stellar
velocity in light units) and are well understood
\citep{gillessenredshift}.  In this paper we have calculated
additional perturbations from general relativity, which are the
following.

\begin{enumerate}
\item The weak-field Schwarzschild effect on the stellar orbit, which
contributes to redshift at $\O(\beta^3)$.  For S2 it is $\simeq
7\kms$.\footnote{Note that these effects are not measurable
separately, only the total redshift is. Hence the values like $7\kms$
depend on the choice of reference orbit and phase, and can change
accordingly.  Nevertheless, the stated numbers give an idea of the
observational precision required.}
\item The frame-dragging effect of black hole spin on the stellar
orbit, which perturbs the redshift at $\O(\beta^4)$.  For S2 it would be
$\sim10^{-1}\kms$ for maximal spin.
\item The weak-field Schwarzschild effect on the light travelling from
the star to us, which gives a redshift perturbation at $\O(\beta^3)$.  For
S2 it is $\simeq 2\kms$.
\item Frame dragging plus next-order Schwarzschild perturbation of the
photon paths.  These contribute at $\O(\beta^5)$ to the redshift, and we
estimate these as $\sim10^{-2}\kms$ for S2.
\end{enumerate}

Of these, the first two are orbital effects and have been considered
in previous work \citep{saha,preto}.  The last two are light-path
effects, known about but not previously computed in the context of
Galactic-center stars.

Clearly, in order to measure the Schwarzschild effects via the
redshift, the effects of general relativity on both the stellar orbit
and the light path must be computed, as they are of the same order.
On the other hand, to leading order, frame dragging needs to be
considered only in the orbit and can be neglected in the light path.

In order to test for the the presence and form of the NLO and NNLO
terms in the metric, the calculated redshift curve must be fitted to
the spectral data via a range of parameters. These include the orbital
parameters and the black hole mass. In keeping these parameters
variable, we can expect a requirement for spectroscopic accuracy less
than the signal sizes themselves. Bear in mind however, that this
depends on the number of data points. The inclusion of astrometric
data to the fitting procedure will help relax the accuracy bound.

Observationally, the Galactic-center stars are of course observable
only in infrared.  For the most massive stars, including S2, the
Brackett-$\gamma$ line at $2.16\,\mu\rm m$ is the most prominent
available spectral feature, and has an intrinsic dispersion of
$\sim100\kms$ \citep[see e.g.,][]{martins}.  For low-mass stars, the
edges of the CO molecular bands (the so-called CO band heads) are
excellent sharp features for redshift measurements.

The SINFONI spectrograph, an instrument at the VLT, has a spectral
resolution of $75\kms$. Redshift errors are currently estimated at 10s
of $\kms$ \citep[see Section 4.1 of][]{gillessen} but expected to
improve.  Measurements by this instrument during S2's next pericenter
passage in 2016 could suffice to provide data from which the
Schwarzschild signals could be extracted.

Spectral measurements seeking the spin-dependent signals are far
beyond the capabilities of existing infrared spectrographs.  On the
other hand, recent developments such as laser-comb spectrographs
\citep[see e.g.,][]{steinmetz} suggest optimism that the spectral
resolutions of next-generation instruments will prove adequate.

Meanwhile, there are some theoretical issues that require further
research.

\begin{itemize}

\item Given a model of gravity which is metric, and an associated
energy-momentum distribution, for which we have a metric, we are able
to calculate the general relativistic redshift as observed by the
Earth. Should we wish to probe the agreement of measured redshift
contributions with a model, an effective way of working backwards
needs to be formulated. Given redshift curve data of sufficient
resolution, methods for determining the model from such must be
developed.

\item The Kerr Black Hole metric is a vacuum solution to the Einstein
field equations. The assumption of a `clean' metric is an
oversimplification.  An extended Newtonian mass distribution in the
galactic center is anticipated. Accreting material, gas, and a
possible accumulation of dark matter in the center could play a
significant role in the dynamics. Such mass distributions need to be
included in the metric.

\item The possibility of a non-Einsteinian black hole metric should
not be disregarded. Alternative theories of gravity possess black hole
solutions whose phenomenology differs potentially already at low order
\citep{will} from weak-field Einstein gravity. Suitable measurements,
combined with the concession of extended mass distributions would
allow for the identification of potentially revealing redshift
contributions. Conclusive results of such studies would likely require
spectral measurements with a resolution beyond that offered by
present-day generation instruments.

\end{itemize}

\acknowledgments
We thank S.~Gillessen and the referee for comments.

\bibliographystyle{apj}
\bibliography{ms.bib}

\end{document}